# Computational Propaganda Theory and Bot Detection System: Critical Literature Review

Manita Pote

## Introduction

According to the classical definition, propaganda is the management of collective attitudes by manipulation of significant symbols. Here, the attitude is not a result of experience but an interference from signs which have a conventionalized significance. Significant symbols are paraphernalia employed in expressing the attitudes or to reaffirm or redefine attitudes. Forms in which significant symbols are embodied to reach the public are spoken, written, pictorial, musical and number of stimulus carriers [1].

With the rise of the internet and social media, the way propaganda is carried out has changed. Automation, bots and human curation are the new means used to distribute misleading information over social media networks to manipulate public opinion, for political polarization etc. This is called computational propaganda. The forms and ways computational propaganda takes place is different from classical propaganda because of the decentralized mode of content proliferation and potential anonymity provided by social media/internet. Hence the classical propaganda theory needs to be revised and redefined to suit the present context. Most of the research on computational propaganda has focused on how it is being carried out in social media but not on the effect on the public opinion and polarization. Thus, there is a need for more research on understanding the effect of computational propaganda.

To detect automation and bots, different machine learning frameworks have been proposed. Most of the frameworks use a supervised learning approach based on the manually labeled data. The problem with these frameworks is the scalability and generalizability. Also these frameworks can identify only a single account as bot or not, not the coordinated activities of botnets. The features used on detection are highly dependent on the data structure made available by social media platforms and mainly focuses on the user specific data and user network. However to make the system more accurate, image related features can also be included. Most of the bot detection system has focused on Twitter because of the data made available by Twitter. However, there are still uncharted areas of research on bot detection systems in other social media platforms along with cross platform activities of bots.

**Propaganda**

Propaganda is the manipulation of collective attitudes of people by using significant symbols embodied in the form of spoken, written, pictorial, musical and number of stimulus carriers [1]. The main aim of propagandists is to multiply all the suggestions favorable to the attitudes which they wish to produce and strengthen and to restrict all suggestions which are unfavorable to them. There are different categories of propaganda like political propaganda (employed by the government), social propaganda, propaganda of agitation (led by the party seeking to destroy the government), self-reproducing propaganda, vertical propaganda (led by leaders), horizontal propaganda (made inside the group) etc [2]. Propaganda can mainly take three forms based on the source and its accuracy - white, grey and black. White propaganda comes from a source that is identified correctly and the information in the message tends to be accurate. When the source is concealed the propaganda is called black propaganda and grey propaganda is somewhere between white and black propaganda [3]. There are different devices to detect propaganda, they are - name calling (propagandists appeal to our hate and fear), glittering generalities (using virtue words like truth, freedom, honor etc.), transfer (carry over of authority we respect), testimonial ( anything using testimonials like patent, policy etc.), plain folks (used by leaders to win our confidence), card stacking (using art of deception) and bandwagon devices (following the crowd) [4].

Digital media platforms introduce new modalities of propaganda such as use of political bots and state-organized 'troll armies' for social astroturfing to simulate public support or opposition towards a particular topic, disguised based propaganda based on cloaked identity created through profile page, stream of user generated content and identity [5]. With the digital age, the forms of propaganda along with the way it is carried out has changed. Now it is called computational propaganda.

Computational propaganda is the use of algorithms, automation and human curation to purposefully distribute misleading information over social media networks to manipulate public opinion, for political polarization etc [6]. With the advent of digital communication and technological change, much of the political communication is taking over social media and political bots - algorithms that operate over social media, written to learn from and mimic real people to manipulate public opinion, are invading political conversation worldwide making up to 50% of all online traffic [7]. In 2016 US election, automated accounts were found to be used deliberately and strategically through out the election, especially by pro-Trump campaigners and programmers carefully adjusted the timing of content production during debates [8]. Russian trolls and bots were also used to spread misinformation and politically biased information in the 2016 US election [9]. Automation associated with anti-Chinese-state perspectives aimed at diasporic chinese were found on Twitter [10]. Computational propaganda is being used all over the world. For example: Woolley, S. C. et. al. [11] found that algorithms, automation and human curation was purposefully used to distribute misleading information over social media networks during elections, political crisis and national security incidents in Brazil, Canada, China, Germany, Poland, Taiwan, Russia, Ukraine and United States.

. The computational propaganda is socially and democratically problematic for mainly three reasons: 1) its production of wrongly informed citizens, 2) likely to stay wrongly informed in echo chambers and 3) be emotionally antagonised or outraged given the effective and provocative nature of fake news [13]. The internet, once supposed to democratize the flow of information, bring more openness and interconnectedness has become a threat to democracy owing to the flawed structural design of the internet itself and the behavior of its users. Digital tools employed in computational propaganda like algorithmic selection (filter bubbles/echo chambers), surveillance (voter targeting/political engineering), political bots and cyber troops are detrimental to democracy [14].

We know understanding of classical propaganda has changed the way it takes place in the digital medium. Hence classical propaganda theory needs to be revised in the digital age. There is no distinction between vertical and horizontal propaganda. Digital media has blurred the line between vertical and horizontal propaganda as individual groups and powerful organizations can all potentially create and orchestrate disguised campaigns within the same online environment. Along with this not only the state actors but foreign entities can also carry out the propaganda, for example Russian trolls in 2016 US election [9], engagement of right-leaning users from Saudi Arabia and Turkey in 2020 US election [15]. Decentralized mode of content proliferation and potential anonymity of social media provide a fertile ground for this.

Racial justice and equality has been the topic of conversation for more than decades. In the current situation, the forms of propaganda previously described - black, grey and white have racist connotations and are offensive to specific communities. Hence politically correct wording needs to be chosen for the forms of propaganda.

Significant symbols used in classical propaganda are different in the digital age. Automation, bots, botnets, web spamming, Twitter bombs, microtargeting, follower padding, conspiracy theories, echo chambers, fake grassroot movements, pro-government or pro-candidate microblog message, misinformation, information laundering are few of the ways propaganda is carried out. In the 2020 US election, conspiratory theories, 'QAnon', 'gate conspiracies' and covid conspiracies, formed a significant portion of discussion on Twitter [15]. Hence, classical propaganda theory needs a new definition.

Most of the research on computational propaganda were found to be on how automation and bots have been used in political issues all over the world. We do not know the effects of digital propaganda, like promotion of echo-chambers of hate and online escalation of political polarization, have on the opinion and behavior of the people exposed to them. Understanding these effects is especially challenging because of the limited types of data social media platforms provide. Social media platforms provides data on user account, its followers, posts and retweets but not emotional nature of misinformation, how often people encounter evocative misinformation, how are individuals exposed to emotional content that is false, to what degree does misinformation play on emotions stemming from ideological, political, racial or religious biases, how frequently do people engage in misinformation etc. Collaboration between social media platforms and academic researchers seems to be necessary to understand the effect.

**Bot detection System**

Social bots are the integral part of computational propaganda. They are social media accounts controlled in part by software. Malicious bots serve as important instruments for orchestrated, large-scale opinion manipulation campaigns on social media. Bots have been actively involved in online discussions of important events, including US elections in 2016 [8] and 2020 [15] as well as in Brexit [16]. Bots have been deployed by powerful elites during political elections to demobilize an opposing party's followers, to target cyber-security threat or political-cultural threats from other states, to attack in-state targets and to send out pro-government or pro-candidate microblog messages [17]. Bots have also been used to alter the number of followers of election candidates, web spamming, send unsolicited replies to specific users via Twitter, to create fake grassroot movements 'astroturf', micro target the online search ads etc [18]. Woolley, S. C. [19] found that network of bots actually played an important role in information flows on Twitter during the 2016 US election campaign by manufacturing consensus or giving the illusion of significant online popularity in order to build real political support and by democratizing online propaganda through enabling nearly anyone to amplify online interactions for partisan ends.

To curb the manipulation and effect of these social bots, it is necessary to identify them and remove from the social network. There are different kinds of social bots and the task of detecting bots is challenging due to their diverse and dynamic behaviors. For example, some bots act autonomously with minimal human intervention, others are manually controlled so that a single entity can create the appearance of multiple human accounts [20]. Some bots are active continuously, others focus bursts of activity on different short-term targets. Different bot-like-behavior are high number of posts in short period of time (usually more than 50), less personal information, more number of retweeting, liking or quoting other users with few or no original posts, large number of accounts retweeting same post at same time to amplify single post, stolen or shared photo, alphanumeric scrambles generated by algorithms as bot name or handle name, highly political posts in multiple languages, use of shortened URLs, same sequence of user liking and retweeting multiple posts in exact same order etc [21].

It is extremely difficult for humans with limited access to social media data to recognize bots, making people vulnerable to manipulation. According to an experimental research among 656 people, participants were able to identify bot accounts with just 71% accuracy among 20 profiles and conservative bots were more likely to be misidentified than liberal bots [22]. There are many machine learning frameworks developed by researchers to detect social bots on Twitter. There are mainly four types of social bot detection systems - graph based, crowd sourced, feature-based and combination of multiple approach systems [23].

One of the most famous systems is **Botometer** developed by Indiana University which uses 1150 features trained using supervised machine learning algorithm, Random Forest algorithm [24]. It uses network structure, user features, friends network, temporal features, linguistic cues and sentiment features. Supervised machine learning algorithms work by training on manually prelabeled data. Botometer provides the range of score to determine whether the twitter account is bot or not by accessing the data of the account provided by Twitter. A score of 0.5 and more is considered a probability of account being a bot. An unsupervised algorithm based framework has also been proposed which uses no human labeled datasets and leverages the similarity in post timelines [25], action sequences [26], content [27] and friends/followers [28]. The disadvantage of unsupervised algorithms is that they are extremely slow. Faster classification can be obtained using fewer features and logistic regression [29][30]. The problem with supervised bot detection systems is that they are not generalizable and scalable as well as it is not possible to detect real-time manipulation campaigns [30]. The reason for this is they are trained on one set of manually labeled data and can not perform well on cross domain data. To overcome these problems, a framework based on just profile information has been proposed which enables to analyze a large-volume stream of accounts in real time [31].

Most of the bot detection systems can identify whether a single account is bot or not. But bots are evolving and becoming much more sophisticated and difficult to detect. Social media platforms themselves are supplying communication resources to bots to escape detection by providing data of popularity measures, automated conversation tools along with vast amounts of user data [32]. Newer bots are more similar to legitimate human-operated accounts and have very less pattern to identify as bots. The reason for the human-like appearance of many bots is an increased hybridization between automated and human driven behaviors which are called cyborgs. Cyborgs operate halfway between the traditional concepts of bots and humans resulting in weakened distinctions and overlapping behaviors between two. There is very little distinction between humans vs bots and fake vs real. Moreover, bots are using AI techniques to generate credible texts and profile pictures [33]. A sophisticated approach that bots have adapted is forming botnets and acting in coordination and synchronization with other bots to amplify their effects while evading the bot detection systems [34].

The major challenge for bot detection systems is the limited amount of data that social media platforms provide through their API service. There is a limit to how much data one can access in a single query which makes it extremely difficult for researchers to collect data. The bot detection system that we currently have is based on the limited data which questions the generalizability of the system. Along with this, the detection system is based on the fixed data structure that Twitter API provides, for example - Botometer. When we provide a user handle to detect whether the account is bot or not, the botometer system requests the data to twitter API and based on the data provided by Twitter, the system provides the score for the account. The system is specifically dependent on the data structure provided by the API. If the changes are made to API data structure, the systems can no longer work. Twitter has recently made available a new API system for academic research with greater data availability which means the current system that is available will need to be migrated to the new system. This can be taken as a lesson for the researchers to make the systems that are independent of data structure of specific social media platforms.

Most of the research has focused on the misinformation and bots for Twitter. The main reason for this is the fact that Twitter makes the data available more easily compared to other social media platforms like Facebook and Youtube. To get the required data from other platforms, researchers need to scrape the data themselves which is a manual and tedious process. However, research on automation and bot detection in other platforms is also necessary. Another uncharted area of research is cross platform activity of bots, for example - sharing of URLs between Facebook and Twitter, Facebook and Instagram etc.

Current bot detection systems can identify only a single account as bot or not but not the coordinated activity of bots. Detecting coordination requires a lot of data as well as the task is computationally challenging as it requires comparing each account. The main problem with adding coordination detection in bot detection systems is the limited data the Twitter makes available. Also because of the computational complexity, the detection will not be real time. Hence, further research is necessary to optimize the algorithm so that the coordination detection can be integrated as well as made real time.

Another unexplored characteristic in bot detection systems is the bots which generate deep fakes and image related misinformation. Microsoft recently developed a software to spot deep fakes, computer manipulated images in which one person's likeness has used to replace that of another [35]. The software analyses photos and videos to give confidence scores about whether the material is likely to have been artificially created. Experts have said that it risks becoming quickly outdated because of the pace at which deepfake tech is advancing [35]. Bots in social media platforms have specifically made use of deepfake technology to generate fake profile pictures as well as to spread misinformation. The current bot detection system we have has not made use of any of the image related characteristics to identify bots. The bot features that have been mainly explored are the text content and friendship structure based on the Natural Language Processing and Network analysis approach. Hence by integrating image processing and image based features, the accuracy of the social bot detection system can be further increased.

## Conclusion

According to the classical definition by Lasswell (1927), propaganda is the management of collective attitudes by manipulation of significant symbols. However this definition has changed to computational propaganda, the way manipulation takes place in digital medium. Computational propaganda is the use of algorithms, automation and human curation to purposefully distribute misleading information over social media networks to manipulate public opinion, for political polarization etc. Digital media platforms have introduced new modalities of propaganda such as the use of social bots and state-organized 'troll armies' for social astroturfing to simulate public support or opposition towards a particular topic. Along with this digital media has blurred the line between different forms of propaganda. Hence existing conceptual and epistemological frameworks in propaganda studies need a revision.

One of the methods to detect the computational propaganda is to identify automation and bots. Many supervised machine learning based frameworks have been proposed for bot detection but these systems can only identify single accounts, not the coordinated activities of botnets and also these systems depend on the data structure provided by the social media platforms. Similarly, current systems have not included the image features in their detection system. Most of the systems are mainly built for Twitter while there are still uncharted areas of research in other social media platforms. Therefore, there are many unexplored research questions and methods in bot detection systems.